\documentclass{aa}
\usepackage{times,graphics}
\begin{document}
\thesaurus{11.01.2; 11.17.3; 13.25.2}
\title{ROSAT HRI observations of seven high redshift quasars}
\author{J.\,Siebert \and W.\,Brinkmann}
\mail{J.\,Siebert, jos@mpe.mpg.de}
\institute{Max-Planck-Institut f\"ur extraterrestrische Physik,
 Giessenbachstrasse, D-85740 Garching, Germany}
\date{Received / Accepted}
\maketitle
\begin{abstract}
We present ROSAT HRI observations of six quasars with redshifts $\ga$3.4:
\object{PKS 0335-122}, \object{S4 0620+389}, \object{B1422+231}, \object{PKS 
2215+02}, \object{Q 2239-386} and \object{PC 2331+0216}. We include the 
observation of the radio-quiet quasar \object{BG 57 9}, whose redshift has 
recently been revised to $z = 0.965$. \object{S4 0620+389}, \object{B1422+231},
\object{PKS 2215+02} and \object{BG 57 9} are detected in the ROSAT energy 
band (0.1-2.4 keV). $2\sigma$ upper limits are given for the remaining objects.
All X-ray sources are point like to the limit of the HRI point spread function.
We report marginal evidence for X-ray variability in \object{S4 0620+389} on 
timescales of several hours (rest frame). No significant X-ray variability is 
found in the gravitationally lensed quasar \object{B1422+231} on rest frame 
timescales of hours and months. We present the spectral energy distributions 
for six of the high redshift quasars and compare them to the mean distribution
of low redshift quasars.
\end{abstract}

\section{Introduction}

The study of quasars at high redshifts is motivated by a number of
interesting, yet unsolved questions: How (and when) have quasars formed?
How do their spectral energy distributions change with time? What causes
the differences between radio-loud and radio-quiet quasars and do these
differences persist to high redshifts? The X-ray emission of quasars is 
particularly interesting, since it constitutes a large fraction of the 
bolometric luminosity of quasars (e.g. Elvis et al. \cite{elvis}). Further, 
the X-rays are mostly produced in the innermost regions of the AGN and 
can therefore provide clues to our understanding of the 'central engine'
and quasar evolution.

In the past, quasars with $z > 3$ have mostly been studied in the optical and 
the radio band. Only the advent of missions like ROSAT (Tr\"umper 
\cite{truemper}) and ASCA (Tanaka et al. \cite{tanaka}), which combine high 
sensitivity with sufficient spatial and spectral resolution, has enabled the 
investigation of the  X-ray properties of these high redshift objects. 
However, the number of quasars with $z >3$, which are detected in X-rays, is 
still low (e.g. Cappi et al. 1997). In order to draw statistically 
significant conclusions on the emission properties of high redshift quasars 
and possible differences to their local counterparts, it is clearly desirable 
to increase the number of sources with X-ray data.

We therefore observed with the ROSAT High Resolution Imager (HRI) four 
radio-loud (\object{PKS 0335-122}, \object{S4 0620+389}, \object{B1422+231}, 
\object{PKS 2215+02}) and two radio-quiet quasars (\object{Q 2239-386}, 
\object{PC 2331+0216}) at $z > 3.4$. The redshift of the radio-quiet quasar 
\object{BG 57 9}, originally cataloged with $z = 3.74$ (Hewitt \& Burbidge 
\cite{hewitt87}), was recently revised to $z = 0.965$ (Borra et al. 
\cite{borra}). Nevertheless, we present the X-ray observation in this paper. 
The basic properties of the observed quasars are summarized in Table 
\ref{prop}. All quasars were previously not detected in X-rays.
Due to its high spatial resolution and the achievable positional accuracy 
the HRI is very well suited to unambiguously identify the X-ray emission of 
high redshift quasars. 

\section{Observations and data reduction}

The X-ray observations were performed with the ROSAT HRI between December 
1994 and August 1996. Details of the individual observations are given in 
Table \ref{log}. 

\begin{table*}
\caption{Properties of the observed high redshift quasars.}
\label{prop}
\begin{tabular}{@{}lclrlrc}
\noalign{\smallskip}
\hline \hline
\noalign{\smallskip}
\multicolumn{1}{c}{Name} & \multicolumn{1}{c}{Position (J2000.0)} & 
\multicolumn{1}{c}{z} & \multicolumn{1}{c}{Ref.} & \multicolumn{1}{c}{m} 
& \multicolumn{1}{c}{Ref.} & $f_\mathrm{5\,GHz}$ \\
\noalign{\smallskip}
\multicolumn{1}{c}{(1)} & \multicolumn{1}{c}{(2)} & \multicolumn{1}{c}{(3)} & 
\multicolumn{1}{c}{(4)} & \multicolumn{1}{c}{(5)} & \multicolumn{1}{c}{(6)} & 
\multicolumn{1}{c}{(7)}\\
\noalign{\smallskip}
\hline
\noalign{\smallskip} 
\object{PKS 0335-122} & 03 37 55.7 $-$12 04 12.0 & 3.442 & O94 & r19.79 & C85 
& 293$\pm$19 mJy\\
\noalign{\smallskip}
\object{S4 0620+389}  & 06 24 19.0 $+$38 56 48.7 & 3.469 & X94 & r20.0 & SK96 
& 836$\pm$74 mJy\\
\noalign{\smallskip}
\object{B1422+231}    & 14 24 38.1 $+$22 56 00.1 & 3.62  & P92 & v16.71 & R93 
& 548$\pm$49 mJy\\
\noalign{\smallskip}
\object{PKS 2215+02}  & 22 17 48.3 $+$02 20 12.1 & 3.581 & S90 & b21.97 & D97 
& 513$\pm$46 mJy\\
\noalign{\smallskip} \hline \noalign{\smallskip}
\object{BG 57 9}      & 13 07 02.7 $+$29 18 42.3 & 3.74  & V82 & v20.14 & B90 &
 ... \\
             &   ...                    & 0.965 & B96  & ...   & ... & 
 ... \\
\noalign{\smallskip}
\object{Q 2239-386}   & 22 41 51.8 $-$38 20 17.2$^{\dagger}$ & 3.511 & ST90 & 
...  & ... & ... \\
\noalign{\smallskip}
\object{PC 2331+0216} & 23 34 32.0 $+$02 33 21.8 & 4.093 & S94 & r19.98 & S89 
& 3.3 mJy$^{\ddagger}$ \\
\noalign{\smallskip}
\hline
\noalign{\smallskip}
\end{tabular}
\vskip 0.2cm
$^{\dagger}$ Uncertainty of the position $\sim 150\arcsec \times 150\arcsec$\\
$^{\ddagger}$ Radio flux from Schneider et al. (1992)\\
(1) Common name; (2) Optical position (J2000.0); (3) Redshift; 
(4) Reference for the redshift: 
Borra et al. \cite{borra} (B96),
Osmer et al. \cite{osmer} (O94), 
Patnaik et al. \cite{patnaik} (P92), 
Savage et al. \cite{savage} (S90), 
Schneider et al. \cite{schneider94} (S94), 
Steidel \cite{steidel} (ST90), 
Vaucher \cite{vaucher} (V82),  
Xu et al. \cite{xu} (X94); 
(5) Optical magnitude. The letter indicates the observing band: 
$b = B_\mathrm{J}$ ($\approx$ 4680{\AA}), $v =$ Johnston V 
($\approx 5500${\AA}), 
$r =$ Gunn r ($\approx$ 6500{\AA}); 
(6) Reference for the optical magnitude: 
Beauchemin et al. \cite{beauchemin} (B90), 
Chu et al. \cite{chu} (C85),
Drinkwater et al. \cite{drinkwater} (D97), 
Remy et al. \cite{remy} (R93), 
Schneider et al. \cite{schneider89} (S89), 
Stickel \& K\"uhr \cite{stickel} (SK96); 
(7) Total 5\,GHz radio flux density from the GB6 radio catalog (Gregory et al. 
\cite{gregory}).    
\end{table*}

The data analysis was performed using standard routines within the EXSAS 
environment (Zimmermann et al. \cite{zimmermann}). The count rates were 
determined with the standard maximum likelihood source detection algorithm,
which delivers the source counts within 2.5 times the FWHM of the HRI point
spread function. In case of no
detection at the optical position of the quasar, i.e. if the respective
likelihood was below 10, which corresponds to about 4$\sigma$, we calculated
the 90\% upper limit for the count rate. The conversion to fluxes in the 
0.1--2.4 keV energy band was done by assuming an average power law spectrum 
for the sources plus Galactic absorption (Dickey \& Lockman \cite{dickey}). 
Since no spectral information is available, the average power law index for 
high redshift quasars was adopted ($\Gamma \sim$ 1.7 and $\Gamma \sim$ 2.0 
for radio-loud and radio-quiet quasars, respectively [Brinkmann, Yuan \& 
Siebert 1997; Yuan et al. 1998]). Finally, luminosities were computed assuming
$H_0 = 50$ km s$^{-1}$ Mpc$^{-1}$, $q_0 = 0.5$ and isotropic emission. The 
results of the analysis and the derived X-ray properties are summarized in 
Table \ref{log}.

\section{Notes on individual sources}

\begin{table*}
\caption{ROSAT observation log and results.}
\label{log}
\begin{tabular}{@{}lcrrrrrr}
\noalign{\smallskip}
\hline \hline
\noalign{\smallskip}
\multicolumn{1}{c}{Name} & \multicolumn{1}{c}{Obs. date} & \multicolumn{1}{c}{Exp.} & 
\multicolumn{1}{c}{cts/s} & \multicolumn{1}{c}{$N_\mathrm{H}$} & 
\multicolumn{1}{c}{$f_\mathrm{x}$} & \multicolumn{1}{c}{$L_\mathrm{x}$} \\
\noalign{\smallskip}
 & & \multicolumn{1}{c}{\small [s]} & \multicolumn{1}{c}{\small 
[$10^{-3}$ s$^{-1}$]} 
 & \multicolumn{1}{c}{\small [cm$^{-2}$]} & 
 \multicolumn{1}{c}{\small [erg cm$^{-2}$ s$^{-1}$]} & 
 \multicolumn{1}{c}{\small [erg s$^{-1}$]} \\
\noalign{\smallskip}
\multicolumn{1}{c}{(1)} & \multicolumn{1}{c}{(2)} & \multicolumn{1}{c}{(3)} & 
\multicolumn{1}{c}{(4)} & \multicolumn{1}{c}{(5)} & \multicolumn{1}{c}{(6)} &
\multicolumn{1}{c}{(7)} \\
\noalign{\smallskip}
\hline
\noalign{\smallskip} 
\object{PKS 0335-122} & 08/27 - 08/28/96 & 1423  & \multicolumn{1}{c}{$<$1.26}
& 4.08 &
$< 7.3\times 10^{-14}$ & $<7.8\times 10^{45}$ \\
\noalign{\smallskip}
\object{S4 0620+389}  & 04/05 - 04/08/96 & 17808 & 6.53$\pm$0.64 & 17.10 &
$(5.8\pm 0.6)\times 10^{-13}$ & $3.5\times 10^{46}$ \\
\noalign{\smallskip}
\object{B1422+231}a   & 01/27/95         & 5678  & 14.98$\pm$1.67 & 2.49 &
($7.7\pm 0.9)\times 10^{-13}$ & $5.1\times 10^{46}$ \\
\noalign{\smallskip}
\object{B1422+231}b   & 01/18 - 01/26/96 & 10918 & 12.43$\pm$1.10 & 2.49 &
($6.4\pm 0.6)\times 10^{-13}$ & $4.2\times 10^{46}$ \\
\noalign{\smallskip}
\object{PKS 2215+02}  & 05/29 - 06/05/95 & 14671 & 5.40$\pm$0.64 & 4.76 & 
($3.2\pm 0.4)\times 10^{-13}$ & $2.1\times 10^{46}$ \\
\noalign{\smallskip}  \hline \noalign{\smallskip}
\object{BG 57 9}        & 07/10 - 07/13/95 & 10630 & 0.93$\pm$0.34 & 2.97 & 
($5.0\pm 1.8)\times 10^{-14}$ & $2.7\times 10^{44}$$^{\ddagger}$ \\
\noalign{\smallskip}
\object{Q 2239-386}   & 05/24 - 05/25/95 & 17094 & \multicolumn{1}{c}{$<$0.38}
 & 
4.69 & $<2.3\times 10^{-14}$ & $<2.3\times 10^{45}$ \\
\noalign{\smallskip}
\object{PC 2331+0216} & 12/24 - 12/26/94 & 15771 & \multicolumn{1}{c}{$<$0.62}
 & 
1.18 & $<2.7\times 10^{-14}$ & $<3.8\times 10^{45}$ \\
\noalign{\smallskip}
\hline
\noalign{\smallskip}
\end{tabular}
\vskip 0.3cm
$^{\ddagger}$ Assuming $z = 3.74$: $L_\mathrm{x} = 5.7\times 10^{45}$ erg 
s$^{-1}$\\
(1) Common name; (2) Date of the HRI observation; (3) Effective, vignetting
and dead time corrected exposure in seconds; (4) Count rate or 2$\sigma$ upper
limit; (5) Galactic $N_\mathrm{H}$ from Dickey \& Lockman (\cite{dickey}) and 
Stark et al. (\cite{stark}); (6) unabsorbed 0.1--2.4 keV flux; (7) 
0.1--2.4 keV luminosity assuming $H_0 = 50$ km s$^{-1}$ Mpc$^{-1}$ and 
$q_0 = 0.5$.
\end{table*}

\subsection{\object{PKS 0335-122}, $z = 3.442$} 

This is an optically faint radio-loud quasar with a Gunn r ($\approx 6500$\AA)
magnitude of 19.79 (Chu et al. \cite{chu}). On the digitized POSS plates a 
stellar objects with $m_\mathrm{B} = 21.44$ is found. The radio spectrum is 
flat ($\alpha^5_{2.7}\approx$0.08, $S_{\nu}\propto \nu^{\alpha}$; 
V\'eron-Cetty \& V\'eron \cite{veron}, VV93 hereafter). The 5\,GHz radio 
emission is clearly variable, since the radio flux density is 293$\pm$19 mJy 
according to the PMN catalog (Griffith \& White \cite{griffith}), whereas the 
higher resolution NVSS (Condon et al. \cite{condon}) gives 475.7$\pm$14.3 mJy.

The exposure of \object{PKS 0335-122} was only $\sim$1400 seconds and the 
source was not detected in this observation. The resulting 2$\sigma$ upper 
limits are $7.3\times 10^{-14}$ erg~cm$^{-2}$~s$^{-1}$ and $7.8\times 10^{45}$ 
erg~s$^{-1}$ for the (unabsorbed) 0.1--2.4 keV flux and luminosity, 
respectively. We can roughly estimate the expected X-ray luminosity by using 
the known correlation with total radio luminosity, e.g. 
$\log L_\mathrm{0.1-2.4 keV} = 0.7 \log L_\mathrm{total, 5\,GHz} + 21.3$ 
(Brinkmann et al. \cite{brinkmann}). It turns out that the upper limit is 
almost a factor of two higher than the estimated X-ray luminosity of 
$\sim 4\times 10^{45}$ erg~s$^{-1}$.  

\subsection{\object{S4 0620+389}, $z = 3.469$}

Optically faint ($m_\mathrm{R} = 20.0$), but radio bright quasar. The total 
5\,GHz flux density is 836$\pm$74 mJy (Gregory et al. \cite{gregory}). In the 
NVSS it had 808.6$\pm$24.3 mJy. \object{S4 0620+389} has been the target of 
various VLBI observations (e.g. Morabito et al. \cite{morabito}; Xu et al. 
\cite{xu}). The radio source was not resolved and it displayed a 5\,GHz VLBI 
flux density of 436 mJy (Xu et al. \cite{xu}). A comparatively high degree of 
polarization of the radio emission has been reported (P$\sim$4\%; Okudaira 
et al. \cite{okudaira}).

\object{S4 0620+389} is a fairly strong X-ray source with an unabsorbed 
0.1--2.4 keV flux of $5.8\times 10^{-13}$ erg~cm$^{-2}$~s$^{-1}$ and a 
corresponding luminosity of $3.5\times 10^{46}$ erg~s$^{-1}$ in the HRI 
observation. In view of this relatively high X-ray flux, we analyzed the 
latest reprocessing of the ROSAT All-Sky Survey (RASS) data for this source. 
We find an X-ray source at the position of \object{S4 0620+389} with 
$\sim 3.5\sigma$ significance. The count rate is 0.023$\pm$0.009, which 
corresponds to ($7.5\pm 2.9)\times 10^{-13}$ erg~cm$^{-2}$~s$^{-1}$. The RASS 
flux is thus fully consistent with the HRI flux.

We compared the radial profile of \object{S4 0620+389} with the properly 
normalized intensity profile of the BL Lac object \object{PKS 2155-304} and 
found no evidence for extended emission to the limit of the point spread 
function of the HRI (FWHM $\sim$5\arcsec, which corresponds to about 34 kpc 
in the rest frame of \object{S4 0620+389} assuming $H_{0} = 50$ km s$^{-1}$ 
Mpc$^{-1}$ and q$_{0} = 0.5$). 

\begin{figure}
\resizebox{\hsize}{!}{\includegraphics{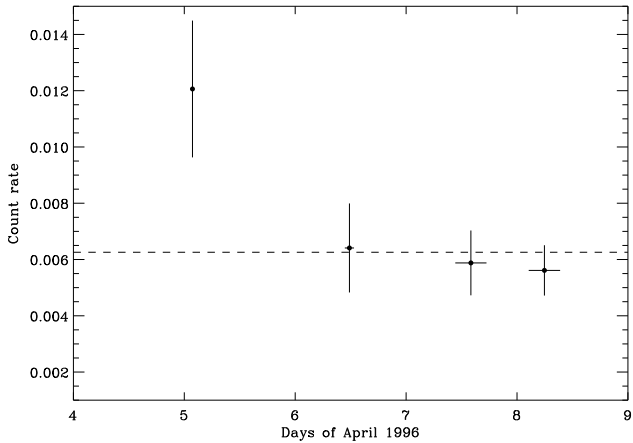}}
\caption{X-ray light curve of \object{S4 0620+389}. The horizontal 
error bars indicate the length of the individual observation interval. 
The dotted line denotes the weighted average of the count rate.}
\label{Slight}
\end{figure}

\object{S4 0620+389} is bright enough to check for variability within the 
observation. We divided the total observation into four contiguous 
observation intervals, one on each day between April, 5 and April, 8 1996,
with individual exposures ranging from 2.2 to 7.8 ksec. We then applied a 
maximum-likelihood source detection on each of these intervals separately. 
The resulting count rates are shown in Fig.~\ref{Slight} as a function of 
time. There is evidence for X-ray variability by a factor of two on a 
timescale of $\sim$1.5 days ($\sim$8 hours rest frame), however, only with 
marginal significance. The highest and the lowest count rate differ by 
$\sim$ 2$\sigma$.
    
\subsection{\object{BG 57 9} (\object{[HB89] 1304+295}), $z = 0.965$}

In the Hewitt \& Burbidge (\cite{hewitt87}; HB87 hereafter) catalog a redshift 
of $z = 3.74$ is given for this source. This value was originally determined 
by Vaucher (\cite{vaucher}). Recently, Borra et al. (\cite{borra}) presented 
new optical data and they report a redshift of $z = 0.965$. Although 
\object{BG 57 9} might therefore not qualify anymore for our high redshift 
sample, we still present its X-ray properties in this paper. The best upper 
limit on the radio emission is f$_\mathrm{1.4\,GHz}< 2.5$ mJy, which 
corresponds to the sensitivity limit of the NVSS. Using $m_\mathrm{v} = 20.14,
f_\mathrm{1.4\,GHz} = 2.5$ mJy and $\alpha_\mathrm{opt} = \alpha_\mathrm{r} = 
0.5$ the K-corrected ratio $R = \log (f_\mathrm{5\,GHz}/f_\mathrm{4400{\AA}}$)
is $\sim$1.5. This is slightly above the commonly accepted dividing line 
between radio-loud and radio-quiet quasars ($R = 1$; Kellerman et al. 
\cite{kellerman}). With present data it therefore cannot be excluded that 
\object{BG 57 9} is a radio-loud quasar. 

\begin{figure}
\resizebox{\hsize}{!}{\includegraphics{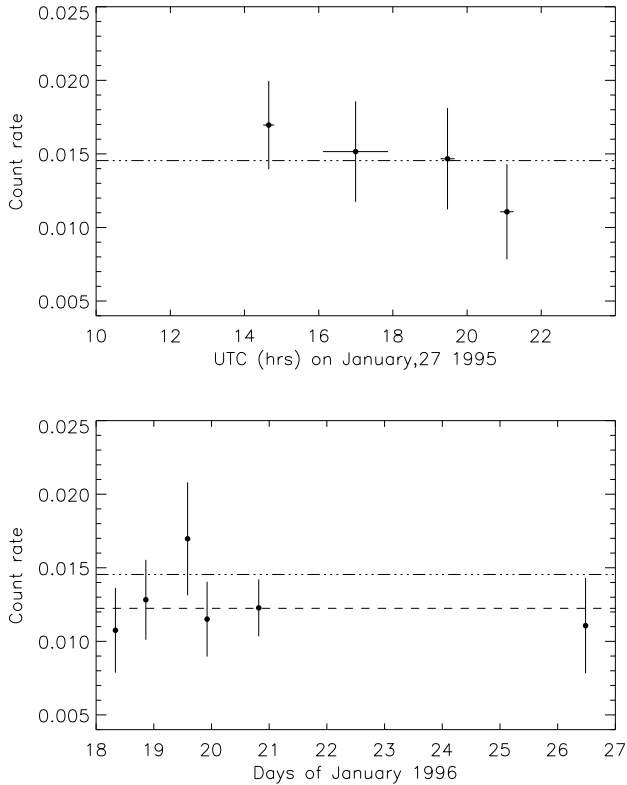}}
\caption{X-ray light curve of the gravitationally lensed system 
\object{B1422+231}. The dash-dotted and the dashed line denote the weighted 
averages of the count rates in the first (top panel) and the second 
observation (bottom panel), respectively.}
\label{Blight}
\end{figure}

The source is marginally detected in our 10.6 ksec HRI observation. The
detection likelihood is 10.6, which corresponds to about 4.3$\sigma$. The
positional difference between the X-ray and the optical position is only 
0.8\arcsec. We are therefore confident that the detected X-ray source is
associated with the quasar. The 0.1--2.4 keV flux of the source is $5.0\times
10^{-14}$ erg~cm$^{-2}$~s$^{-1}$. Applying the new redshift of 
\object{BG 57 9}, this gives a luminosity of $2.7\times 10^{44}$ erg~s$^{-1}$.

\subsection{\object{B1422+231}, $z = 3.62$}

This object is a optically and radio bright, gravitationally lensed quasar at 
a redshift of $z = 3.62$. The original radio map shows four components within 
1.3\arcsec (Patnaik et al. \cite{patnaik}). Subsequent optical and infrared 
observations confirmed the lensing hypo\-thesis and discovered variability in 
the three brightest components (Lawrence et al. \cite{lawrence}; Yee \& 
Bechtold \cite{yee}). A recent HST observation revealed the lensing 
galaxy at a redshift of $z\sim 0.4$ close to the faintest quasar image
(Impey et al. \cite{impey}).

\object{B1422+231} was observed twice with the HRI: for $\sim$5.7 ksec in 
January 1996 and for another $\sim$10.9 ksec about one year later. It is 
clearly detected in both observations and the 0.1--2.4 fluxes are 
$(7.7\pm 0.9)\times 10^{-13}$ erg~cm$^{-2}$~s$^{-1}$ and $(6.4\pm 0.6)\times
10^{-13}$ erg~cm$^{-2}$~s$^{-1}$, respectively. The X-ray emission has not 
varied significantly between the two observations. Analyzing the reprocessed 
ROSAT All-Sky Survey data, we find that \object{B1422+231} is marginally 
detected (3.7$\sigma$). The X-ray flux in the Survey is $(4.5\pm 1.2)\times 
10^{-13}$ erg~cm$^{-2}$~s$^{-1}$, which might indicate that \object{B1422+231}
has varied in X-rays between the All-Sky Survey and the HRI observation
($\Delta T \sim 5$ years, i.e. $\sim 1$ year rest frame). 

Next we looked for X-ray variability within each of the two observations of
\object{B1422+231}. We divided the observations into several contiguous 
intervals with individual exposures between 1.2 and 2 ksec and applied a 
maximum-likelihood source detection on each of the intervals separately. 
The resulting count rates are shown in Fig.~\ref{Blight} as a function of 
time. Although the light curve suggests some trends in the count rate, a
$\chi^2$ test gives no indication for significant variability ($\chi^2_{\nu} 
= 0.58$ (9 dof)).  

\begin{figure}
\resizebox{\hsize}{!}{\includegraphics{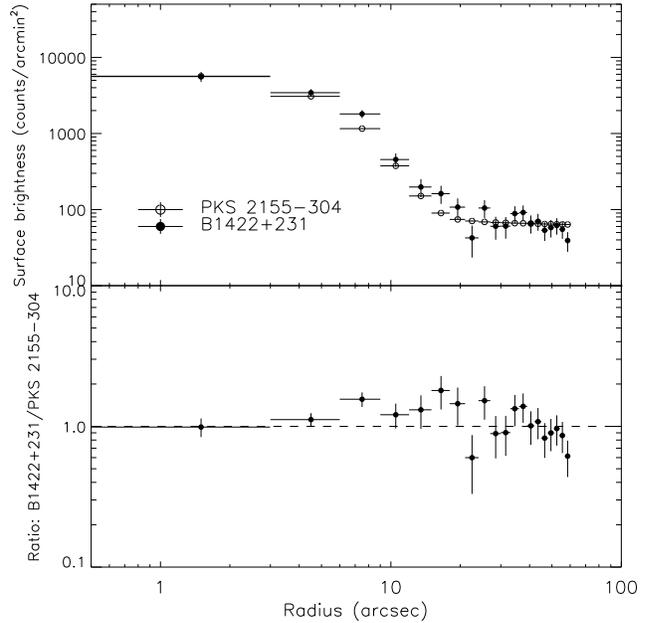}}
\caption{The radial X-ray surface brightness profile of 
\object{B1422+231} compared to that of the BL Lac object PKS 2155-304. The 
latter has been normalized to the first bin of the \object{B1422+231} profile.
The bottom panel shows the ratio of the two intensity profiles. No 
significantly extended X-ray emission is found.}
\label{Brad}
\end{figure}

The X-ray surface brightness profile for \object{B1422+231} is shown in 
Fig.~\ref{Brad} in comparison to that of a $\sim$50 ksec HRI observation of 
the BL Lac object \object{PKS 2155-304}. The profile of \object{PKS 2155-304} 
was normalized to the first bin of the radial profile of \object{B1422+231}. 
The local background intensity was determined from an annulus with inner 
radius 2.5\arcmin\ and outer radius 5\arcmin\ around \object{B1422+231} and 
then added to the intensity profile of \object{PKS 2155-304}. There is no 
evidence for significantly extended X-ray emission to the limit of the PSF of 
the ROSAT HRI (FWHM $\sim$5\arcsec, which corresponds to about 34 kpc for 
$z = 3.62$). The systematic trend, which is visible in the lower panel of 
Fig.~\ref{Brad} up to a radius of 20\arcsec, is most likely due to the energy 
dependent PSF of the HRI, which is slightly more extended for higher energies.
Since the average soft X-ray spectrum of high redshift radio-loud quasars is 
much harder ($\Gamma \approx 1.7$; e.g. Siebert et al. \cite{siebert}) than 
that of \object{PKS 2155-304} ($\Gamma \approx 2.65$; Brinkmann et al. 
\cite{brinkmann94}), the radial profile of \object{B1422+231} appears to be 
slightly extended. 

\subsection{\object{PKS 2215+02}, $z = 3.581$}

Optically very faint ($m_\mathrm{B} = 21.97$) radio-loud quasar at a redshift 
of $z = 3.581$. The radio spectrum of \object{PKS 2215+02} between 2.7\,GHz 
and 5\,GHz is flat ($\alpha = -0.05$, VV93) and the source seems to be 
variable in the radio band, since the flux density from the 87GB survey 
($f_\mathrm{5\,GHz} = 513\pm 46$ mJy; Gregory et al. \cite{gregory}) is 
significantly lower than the flux density measured four years later in the 
PMN survey ($f_\mathrm{5\,GHz} = 642\pm 35$ mJy; Griffith \& White 
\cite{griffith}) with an angular resolution comparable to the 87GB survey. 

The source was clearly detected in our 14.7 ksec HRI observation. The 
X-ray position from the maximum-likelihood algorithm and the best known 
optical position of \object{PKS 2215+02} differ by 4.5\arcsec. This difference
is easily explained by the expected residual boresight errors (e.g. Voges 
1992). The unabsorbed 0.1--2.4 keV flux is 
$3.2\times 10^{-13}$ erg~cm$^{-2}$~s$^{-1}$, which gives an X-ray luminosity 
in the ROSAT band of $2.1\times 10^{46}$ erg~s$^{-1}$. The radial profile is 
consistent with the point spread function of the HRI, therefore no evidence 
for extended X-ray emission is found. 

No source is detected in the RASS and the 2$\sigma$ upper limit 
$(6.4\times 10^{-13}$ erg~cm$^{-2}$~s$^{-1}$) is well above the actual 
X-ray flux measured with the HRI.

\subsection{\object{Q 2239-386}, $z = 3.511$} 

The position of this quasar is only poorly known and the positions given 
in VV93 and Hewitt \& Burbidge (\cite{hewitt93}; HB93) are inconsistent. We 
adopt the position given in HB93 in this paper. The positional uncertainty 
listed in HB93 is of the order of $\sim$150\arcsec. To our knowledge, no 
optical magnitude for this quasar is reported in the literature. Due to the 
poorly constrained optical position, the search for an optical counterpart on 
digitized POSS plates is not feasible (there are about 50 objects within 
150\arcsec\ of the given optical position). Imaging spectroscopy of this field 
is clearly warranted. 

The upper limit to the X-ray flux at the position given in HB93 is
$2.3\times 10^{-14}$ erg~cm$^{-2}$~s$^{-1}$, which corresponds to 
$L_\mathrm{0.1-2.4 keV} = 2.3\times 10^{45}$ erg~s$^{-1}$. 

We note that there is an X-ray source at a distance of $\sim$3\arcmin\ from the
position of HB93 with a 0.1--2.4 keV flux of $\sim 2\times 10^{-13}$ 
erg~cm$^{-2}$~s$^{-1}$. In the vicinity of this X-ray source two objects can 
be found on the digitized POSS plate: a 14.78 mag stellar object at a distance
of 3.3\arcsec\ and a 18.39 mag galaxy at a distance of 9.1\arcsec\ from the 
X-ray position. Both possible identifications suggest that this X-ray source 
is not associated with the high redshift quasar. 

\subsection{\object{PC 2331+0216}, $z = 4.093$} 

This quasar, originally classified as radio-quiet, has the highest redshift 
of the present sample. Schneider et al. (\cite{schneider92}) report a 5\,GHz 
flux density of 3.3 mJy. Using $m_\mathrm{r} = 19.98$, we get $R \approx 1.8$ 
for the K-corrected logarithmic ratio of 5\,GHz radio to 4400{\AA} optical 
flux. Therefore \object{PC 2331+0216} is at the borderline between radio-loud 
and radio-quiet quasars.

\object{PC 2331+0216} was not detected in our $\sim$16 ksec observation with 
the HRI. The 2$\sigma$ upper limit to the 0.1--2.4 keV flux is $2.7\times 
10^{-14}$ erg~cm$^{-2}$~s$^{-1}$, which leads to a luminosity upper limit of 
$3.8\times 10^{45}$ erg~s$^{-1}$. 

\section{Spectral energy distributions}

\begin{figure*}
\resizebox{\hsize}{!}{\includegraphics{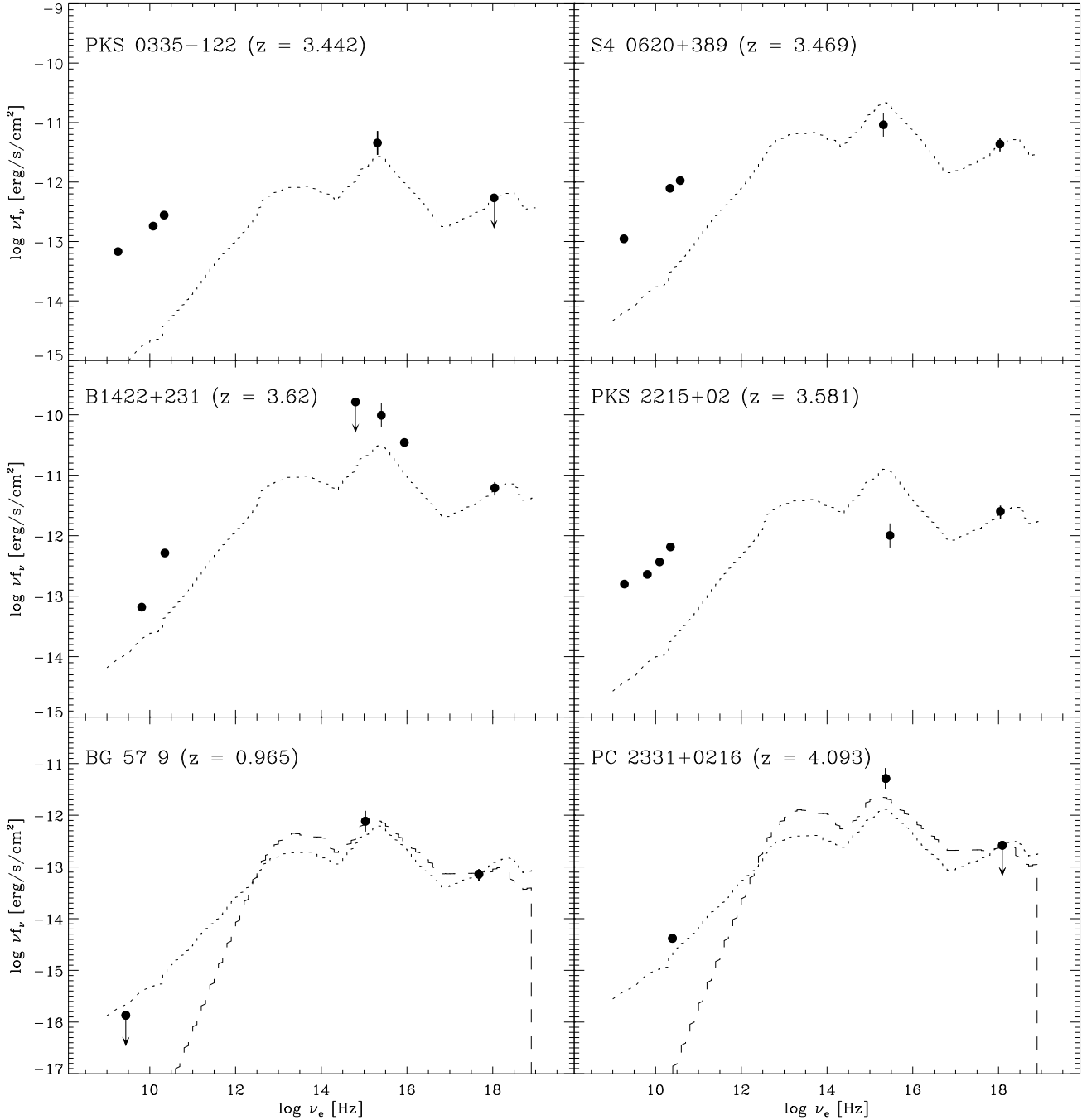}}
\caption{The spectral energy distribution for the four radio-loud quasars 
(upper four panels) and two of the radio-quiet quasars. The dotted and the 
dashed lines represent the mean spectral energy distribution of radio-loud and
radio-quiet quasars, repectively,  as given in Elvis et al. (\cite{elvis}). 
The spectral energy distributions are arbitrarily normalized to the 1 keV 
flux.}
\label{nufnu}
\end{figure*}

In Fig.~\ref{nufnu} we present the spectral energy distributions (SED) from
radio to X-ray frequencies for six of the high redshift quasars. 
\object{Q 2239-386} is omitted from this analysis. All fluxes are given in the
rest frame of the sources. Radio fluxes at various frequencies are used as 
listed in NED. In the case of \object{BG 57 9} we also used the upper limit 
from the NVSS (Condon et al. \cite{condon}). Optical fluxes are calculated 
from the magnitudes given in Table \ref{prop} and corrected for Galactic 
reddening as follows (e.g. Zombeck \cite{zombeck}):
\[
f_{6500} = 10^{-0.4\cdot(m_\mathrm{r} - A_\mathrm{r}) - 19.547} \quad 
\mathrm{erg~cm}^{-2}\, \mathrm{s}^{-1}\, \mathrm{Hz}^{-1}
\]
\[
f_{5500} = 10^{-0.4\cdot(m_\mathrm{v} - A_\mathrm{v}) - 19.435} \quad 
\mathrm{erg~cm}^{-2}\, \mathrm{s}^{-1}\, \mathrm{Hz}^{-1}
\]
\[
f_{4400} = 10^{-0.4\cdot(m_\mathrm{b} - A_\mathrm{b}) - 19.369} \quad 
\mathrm{erg~cm}^{-2}\, \mathrm{s}^{-1}\, \mathrm{Hz}^{-1} 
\]

The Galactic extinction at various wavelengths has been calculated from 
$A_\mathrm{B}$ as given in NED by using the reddening law from Seaton et al. 
(\cite{seaton}). In the case of \object{B1422+231} we added a data point 
in the UV from the IUE observation ($f_{\lambda} \approx 1.1\times 10^{-15}$ 
erg s$^{-1}$ cm$^{-2}$ \AA$^{-1}$ between 1250 and 2000 \AA) and in
the near-infrared from Lawrence et al. (\cite{lawrence}), who give an upper 
limit to the K-band (2.2$\mu$m) emission of 12.7 mag. This magnitude was 
converted to a flux according to the formula (e.g. Wamsteker et al. 
\cite{wamsteker}):

\[
f_{2.2} = 10^{-0.4\cdot m_\mathrm{K} - 20.172} \quad 
\mathrm{erg~cm}^{-2}\, \mathrm{s}^{-1}\, \mathrm{Hz}^{-1}
\] 

The monochromatic X-ray fluxes (and upper limits) at 1 keV were calculated 
assuming $\Gamma = 1.7$ for radio-loud and $\Gamma = 2$ for radio-quiet
quasars. All fluxes were finally transformed to the rest frame of the source.

For comparison we plot in Fig.~\ref{nufnu} the mean SED of radio-loud 
(dotted) and radio-quiet (dashed) quasars as given in Elvis et al. 
(\cite{elvis}). The SEDs are arbitrarily normalized to the X-ray fluxes. 
Since the Elvis et al. (\cite{elvis}) sample contains X-ray bright 
{\it Einstein} IPC detected quasars, the mean SED tends to fall below the 
observed quasars at optical and radio wavelengths when it is normalized to 
the X-ray flux. 

Taking this into account, most of the high redshift SEDs agree reasonably 
well with the low redshift mean. The systematically higher radio emission
is a general feature of the SEDs of high redshift radio-loud quasars (Bechtold
et al. \cite{bechtold}; Siebert et al. \cite{siebert}). Still, remarkable 
differences between the radio-loud quasars are obvious from Fig.~\ref{nufnu} 
with the two extremes being \object{B1422+231} and \object{PKS 2215+02}. The 
main difference appears in the optical, where \object{PKS 2215+02} is about 
two orders of magnitude fainter than \object{B1422+231}, whereas the X-ray 
and the radio fluxes are roughly comparable. This result might be explained 
by significant dust extinction intrinsic to the source. We note, however, 
that Francis et al. (in prep.) give an optical slope of 
$\alpha_{\rm opt}\approx -1$ for \object{PKS 2215+02}, which does not indicate
a particularly red continuum compared to the other flat-spectrum quasars of
their sample. 

The SED of \object{PC 2331+0216} fits remarkably well to the mean SED of low 
redshift radio-loud quasars. In particular the radio flux is almost three 
orders of magnitude higher than expected for an average radio-quiet quasar. 
Therefore \object{PC 2331+0216} should indeed be considered as a radio-loud 
quasar.

In Table \ref{alpha} we list the two-point spectral indices $\alpha_{\rm ox}$,
$\alpha_{\rm ro}$ and $\alpha_{\rm rx}$ between the radio (4.85\,GHz), 
the UV/optical (2500\,\AA) and the X-ray band (2\,keV). The spectral indices
are defined as follows:
\begin{eqnarray*}
\alpha_\mathrm{ox} = 0.384 \times \log (f_\mathrm{o}/f_\mathrm{x}) \\
\alpha_\mathrm{ro} = 0.185 \times \log (f_\mathrm{r}/f_\mathrm{o}) \\
\alpha_\mathrm{rx} = 0.125 \times \log (f_\mathrm{r}/f_\mathrm{x})
\end{eqnarray*}
where $f_\mathrm{x}$, $f_\mathrm{o}$ and $f_\mathrm{r}$ are the rest frame 
fluxes at the three frequencies given above.
 
$\alpha_{\rm ox}$ and $\alpha_{\rm ro}$ both show a large scatter for the
radio-loud quasars and the corresponding flux/luminosity ratios vary by 
factors of 40 to 80. This scatter is obviously mostly due to the optical 
emission, because the radio-to-X-ray spectral index $\alpha_{\rm rx}$ is 
almost constant for all four objects. The $\alpha_{\rm ox}$ values are
roughly consistent with the previously found mean values for larger samples.
Brinkmann et al. (\cite{brinkmann}) give $\langle\alpha_{\rm ox}\rangle = 
1.24$ for a sample of 297 flat-spectrum radio-loud quasars and Bechtold et al.
(\cite{bechtold}) find $\alpha_{\rm ox}\sim 1.4$ for high redshift quasars. 
As already noted, \object{PKS 2215+02} is underluminous in the optical, which 
results in a very flat $\alpha_{\rm ox}$ of 0.9. There is only one high 
redshift quasar known, which exhibits a similarly extreme optical-to-X-ray 
ratio, namely 1745+624 ($\alpha_{\rm ox}\approx 0.8$; Fink \& Briel 
\cite{fink}). 
 
\begin{table}
\caption{Two point spectral indices.}
\label{alpha}
\begin{tabular}{@{}lrrr}
\noalign{\smallskip}
\hline \hline
\noalign{\smallskip}
\multicolumn{1}{c}{Name} & \multicolumn{1}{c}{$\alpha_{\rm ox}$} & 
\multicolumn{1}{c}{$\alpha_{\rm ro}$} & \multicolumn{1}{c}{$\alpha_{\rm rx}$}\\
\noalign{\smallskip}
\hline 
\noalign{\smallskip}
\object{PKS 0335-122} & $>$1.38 & 0.70 & $>$0.92\\
\object{S4 0620+389}  &    1.14 & 0.73 & 0.86\\
\object{B1422+231}    &    1.50 & 0.50 & 0.82\\
\object{PKS 2215+02}  &    0.90 & 0.85 & 0.87\\
\object{BG 57 9}        &    1.42 & $<$0.35 & $<$0.70\\
\object{PC 2331+0216} & $>$1.42 & 0.34 & $>$0.70\\       
\noalign{\smallskip}
\hline 
\noalign{\smallskip}
\end{tabular}
\end{table}

\section{Summary}

We presented ROSAT HRI observations of six quasars with redshifts $z\ga 3.4$
and one with $z = 0.965$. Four objects (\object{S4 0620+389},
\object{B1422+231}, \object{PKS 2215+02}, \object{BG 57 9}) are detected with 
X-ray luminosities ranging from $2.7 \times 10^{44}$ erg~s$^{-1}$ to 
$5.1\times 10^{46}$ erg~s$^{-1}$. 2$\sigma$ upper limits are given for the 
remaining three objects (\object{PKS 0335-122}, \object{Q 2239-386}, 
\object{PC 2331+0216}).

All X-ray sources are point like to the limit of the ROSAT HRI point spread 
function ($\approx$5\arcsec, which corresponds to about 34 kpc at $z = 3.5$).

We find marginal evidence for X-ray variability by a factor of two within the
observation of \object{S4 0620+389}, i.e. on a timescale of several hours in 
the rest frame of the source. The gravitationally lensed system 
\object{B1422+231} shows no indication for X-ray variability during the 
individual observations as well as between the two HRI observations, which 
are separated by about one year.

The spectral energy distributions of six of the high redshift quasars are
presented. Although they are roughly consistent with the mean 
SED of low redshift quasars (Elvis et al. \cite{elvis}), pronounced 
differences between individual objects appear. In particular 
\object{PKS 2215+02} is almost two orders of magnitudes fainter in the 
optical compared to \object{B1422+231}, while the radio and the X-ray 
luminosities differ by only a factor of two. However, the optical continuum 
slope measured by Francis et al. (in prep.) does not indicate significant 
extinction by dust.      

\begin{acknowledgements}
This research has made use of the NASA/IPAC Extragalactic Data Base
(NED) which is operated by the Jet Propulsion Laboratory, California
Institute of Technology, under contract with the National Aeronautics
and Space Administration.
\end{acknowledgements}

\end{document}